\documentclass[AMA,STIX2COL]{MRM}

\articletype{Rapid Communication}%

\received{29 March 2024}
\revised{30 October 2024}
\accepted{26 November 2024}

\usepackage{titlesec}
\usepackage{makecell}
\usepackage{dsfont}
\usepackage{anyfontsize}
\usepackage{xr}
\usepackage{xr-hyper}
\usepackage{float}
\usepackage{multirow}
\usepackage{colortbl}       % Adds color support for tables
\definecolor{customRed}{HTML}{ea4a6b}
\definecolor{customYellow}{HTML}{fdae86}

\usepackage{subfiles} % Best loaded last in the preamble

\AtBeginDocument{%
}

\newcommand{\rev}[1]{\textcolor{black}{#1}}

\makeatletter
\newcommand*{\addFileDependency}[1]{% argument=file name and extension
\typeout{(#1)}% latexmk will find this if $recorder=0
\@addtofilelist{#1}
\IfFileExists{#1}{}{\typeout{No file #1.}}
}\makeatother

\begin{document}

\title{WALINET: A water and lipid identification convolutional Neural Network for nuisance signal removal in ${}^1$H MR Spectroscopic Imaging}

\author[1,2,3]{Paul Weiser}{}
\author[3]{Georg Langs}{}
\author[4]{Stanislav Motyka}{}
\author[4]{Wolfgang Bogner}{}
\author[6,7]{Sébastien Courvoisier}{}
\author[1,2]{Malte Hoffmann}{}
\author[5]{Antoine Klauser}{\textsuperscript{†}}
\author[1,2]{Ovidiu C. Andronesi}{\textsuperscript{†}}

\address[1]{\orgdiv{Athinoula A. Martinos Center for Biomedical Imaging}, \orgname{Massachusetts General Hospital}, \orgaddress{\city{Boston}, \state{MA}, \country{USA}}}
\address[2]{\orgdiv{Department of Radiology}, \orgname{Massachusetts General Hospital, Harvard Medical School}, \orgaddress{\city{Boston}, \state{MA}, \country{USA}}}
\address[3]{\orgdiv{Computational Imaging Research Lab - Department of Biomedical Imaging and Image-guided Therapy}, \orgname{Medical University of Vienna}, \orgaddress{\state{Vienna}, \country{Austria}}}
\address[4]{\orgdiv{High Field MR Center - Department of Biomedical Imaging and Image‐Guided Therapy}, \orgname{Medical University of Vienna}, \orgaddress{\state{Vienna}, \country{Austria}}}

\address[5]{\orgdiv{Advanced Clinical Imaging Technology}, \orgname{Siemens Healthineers International AG}, \orgaddress{\city{Lausanne}, \country{Switzerland}}}
\address[6]{\orgdiv{Center for Biomedical Imaging (CIBM)}, \orgaddress{\city{Geneva}, \country{Switzerland}}}
\address[7]{\orgdiv{Department of Radiology and Medical Informatics, University of Geneva}, \orgaddress{\city{Geneva}, \country{Switzerland}}}
\address[*]{\orgdiv{\textbf{{†}}: These authors contributed equally to this work.}}

%%==================================%%
%% Abstract %%
%%==================================%%

\abstract{
\textbf{Purpose.} Proton Magnetic Resonance Spectroscopic Imaging (${}^1$H-MRSI) provides non-invasive spectral-spatial mapping of metabolism. However, long-standing problems in whole-brain ${}^1$H-MRSI are spectral overlap of metabolite peaks with large lipid signal from scalp, and overwhelming water signal that distorts spectra. Fast and effective methods are needed for high-resolution ${}^1$H-MRSI to accurately remove lipid and water signals while preserving the metabolite signal. The potential of supervised neural networks for this task remains unexplored, despite their success for other MRSI processing.

\textbf{Methods.}  We introduce a deep-learning method based on a modified Y-NET network for water and lipid removal in whole-brain ${}^1$H-MRSI. The WALINET (WAter and LIpid neural NETwork) was compared to conventional methods such as the state-of-the-art lipid L2 regularization and Hankel-Lanczos singular value decomposition (HLSVD) water suppression. Methods were evaluated on simulated and \textit{in-vivo} whole-brain MRSI using NMRSE, SNR, CRLB, and FWHM metrics. 

\textbf{Results.} WALINET is significantly faster and needs 8s for high-resolution whole-brain MRSI, compared to 42 minutes for conventional HLSVD+L2. \rev{WALINET suppresses lipid and water in the brain by 25-45 and 34-53 fold, respectively.} WALINET has better performance than HLSVD+L2, providing: 1) more lipid removal with 41\text{\%} lower NRMSE, 2) better metabolite signal preservation with 71\text{\%} lower NRMSE in simulated data, 155\text{\%} higher SNR and 50\text{\%} lower CRLB in in-vivo data. Metabolic maps obtained by WALINET in healthy subjects and patients show better gray/white-matter contrast with more visible structural details.

\textbf{Conclusions.} WALINET has superior performance for nuisance signal removal and metabolite quantification on whole-brain ${}^1$H-MRSI compared to conventional state-of-the-art techniques. This represents a new application of deep-learning for MRSI processing, with potential for automated high-throughput workflow.
 
}


\wordcount{3665}

\maketitle
%\footnotetext{\textbf{Abbreviations:}~\hbox{ANA,~anti-nuclear~antibodies;~APC,~antigen-}{\hfill\break}presenting~cells; IRF, interferon regulatory factor}

%%==================================%%
%% Introduction %%
%%==================================%%

\section{Introduction}\label{sec0}

%
%\begin{figure}
%  \begin{center}
%    \includegraphics[width=0.5\textwidth]{figures/teaser.png}
%  \end{center}
%  \caption{caption}
%\end{figure}

Proton magnetic resonance spectroscopic imaging (${}^1$H-MRSI) has great potential as a metabolic imaging technique that can measure the intrinsic metabolite concentrations across the whole-brain without the need to administer molecular agents \cite{bogner2021accelerated}. Proton-based MR spectroscopy, compared to X-nuclei spectroscopy, has a significantly higher SNR and is ubiquitously available on all MRI scanners equipped with standard hardware and software. Consequently, it represents the vast majority of MRS performed in clinical studies and research applications. In specific pathology it allows the detection of metabolic abnormalities in the brain before anatomical lesions are visible \cite{oz2014clinical}. Therefore, it is valuable for disease investigation such as brain tumor identification, classification and treatment response assessment \cite{kickingereder2018radiomics}. It further enables the study of neurochemistry alterations and disease mechanisms in neuropsychiatry \cite{su2016whole}. However, the clinical potential of ${}^1$H-MRSI is not fully realized due to complex experimental factors that reduce data quality such as large artifacts from the overwhelming water and lipid signals \cite{maudsley2021advanced, wilson2019methodological}. Hence, efficient solutions to these technical problems can have great and widespread impact for the use of ${}^1$H-MRSI in clinical applications. 

% water signal contamination is present in a whole brain. suppressed during the scan and after. amplitude is orders of magnitude greater and lipids and metabolites

Since water is a major component of all brain tissues (70\% - 85\%), the signal strength of water artifacts is particularly large, resulting in amplitudes 3-4 orders of magnitude greater than those of metabolites. Although the water peak at ~4.68 ppm does not directly overlap with the peaks of the major metabolites in the aliphatic region (~0.9-4.2 ppm), frequency modulation due to timing of the pulse sequence acquisition can lead to large water side-bands that overlap metabolite peaks and cause significant baseline distortions. Additionally, the water suppression pulses change the shape of the water peak from an absorption symmetric peak shape to an asymmetric peak shape where the tails are larger than the center of peak.

The removal of lipid signal originating from the scalp region presents with even greater challenges \cite{tkavc2021water}. The lipid spectrum is complicated, having multiple peaks that fully overlap with aliphatic region and are 1-2 orders of magnitude stronger than metabolite peaks. Furthermore, the lipid signal originates from scalp areas with very inhomogeneous $B_0$ field, hence they are much broader than water and metabolite peaks from inside the brain.
% water suppression methods explain here...
% During acquisition: WET, VAPOR
% After aquisition: HLSVD/HLSVD, Löwner tensor-based blind source separation, L2-regularization

Methods enabling suppression of water and lipid contamination can be categorized into several groups \cite{tkavc2021water}: techniques leveraging specific RF-pulse or sequence designs to invert, nullify or saturate water or lipid resonances \cite{balchandani2008fat, zhu2010dual, hangel2015lipid, ogg1994wet, tkavc1999vivo, esmaeili2017three, weng2022slow, chang2018non} approaches utilizing dedicated hardware to spoil the scalp signal \cite{strasser2022improving, de2018elliptical}, and post-acquisition methods employing spatial or spectral priors for lipid contamination removal.
The latter category in the case of lipids removal can be divided in methods using lipid signal extrapolation \cite{haupt1996removal}, dual-density reconstruction \cite{metzger1999hybrid, sarkar2002truncation}, lipid-basis penalty \cite{bilgic2013lipid, lee2010iterative}, subspace reconstruction methods based on specific spatial supports and spectral decomposition \cite{ma2016removal, adany2021method, bhattacharya2017compartmentalized}, while in the case of water removal includes Hankel matrix singular value decomposition \cite{barkhuijsen1987improved}, subspace reconstruction \cite{ma2016removal}, Lowner tensorization \cite{nagaraja2018tensor}, and water-basis penalty \cite{lin2019water}.

In practice, whole-brain ${}^1$H-MRSI has residual water and lipid signals present even after using special pulse sequences and hardware designed to suppress them, hence requiring post-acquisition processing methods. Optimized processing methods for nuisance signal removal are particularly relevant for ${}^1$H-MRSI at 7T, because $B_0$, $B_1+$  inhomogeneity and high SAR at ultra-high field make lipid and water suppression during acquisition prohibitive and inefficient. 

Due to fundamentally different spatial distribution, spectral ranges and signal shapes of water and lipids, existing postprocessing methods usually only allow the removal of water \textit{or} lipid signals, but not both in the same time. 

Recently, closed form solutions have been developed for lipid removal \cite{bilgic2014fast, tsai2019reduction}. Closed form solutions are derived by applying a linear operator on each individual spectrum and are faster compared to iterative lipid suppression algorithms. However, the linear operator method has some drawbacks too: 1) it is subject specific, 2) requires accurate anatomical mask delineation, and 3) may not work if the orthogonal assumption and linear superposition between lipids and metabolites is not met.. 

Hence, long processing times and the requirement for tedious subject-specific parameter optimizations such as the L2 regularization factor to balance nuisance signal suppression with the preservation of metabolic signals, make difficult the application of existing nuisance signal removal methods for fast, robust and automated MRSI processing pipelines.

In recent years, deep learning-based methods applied to mitigate MRS challenges have gained popularity, enabling robust applications and eliminating the need for complex parameter optimizations. Thereby, applications can be categorized into deep learning based artifact removal, denoising, lowrank \cite{kreis2018artifactcnn, lam2019constrained, lee2019intact} and spectral quantification \cite{gurbani2019incorporation, hatami2018magnetic, louis2021quantification, lee2020deep, starcuk2023quantcnn}. 

Motivated by these developments, we introduce a convolutional neural network for the identification of water \& lipid (WALINET: WAter and LIpid neural NETwork) signals in ${}^1$H-MRSI spectra, and evaluated its performance in simulations and \textit{in-vivo} data measured in human participants.

%Recently, closed form solutions have been developed for these methods \cite{bilgic2014fast, tsai2019reduction}. Thereby, making them more robust and faster compared to iterative lipid suppression algorithms. Closed form solutions are derived by applying a linear operator on each individual spectrum . Its application enables a fast and efficient lipid removal, but reveals also the drawbacks. The linear operator is subject specific and needs to be recomputed for each subject. It is therefore sensitive to potentially inaccurate anatomical mask delineation that would affect its effectiveness. Second, a linear operator is by nature,limited  to represent exclusively linear functions. This inherently prevent any non-linear behavior of the lipid suppression that may improve accuracy of the lipid suppression.
%In this work we present a  convolutional neural network for lipid signal identification and suppression  (LIPNET). L
% was 
%%==================================%%
%% Methods %%
%%==================================%%

\section{Methods}\label{sec1}

\paragraph{Strategy:}
 The problem can be formulated using two distinct inputs $x_1$ and $x_2$: 1) the original MRS spectrum ($x_1$) containing metabolite signal $m$ contaminated with lipid $l$ and water $w$ signal, and 2) the spectrum ($x_2$) subjected to a projection onto the lipid subspace, $\mathds{1}-\mathcal{L}$, using lipid L2 regularization \cite{bilgic2014fast} approach with $\mathcal{L} = (\mathds{1}+\beta\mathbf{LL^H})^{-1}$. $\mathbf{L}$ is a matrix containing the in-vivo lipid signal obtained from the scalp mask of each subject and $\beta$ is the regularization parameter. The derived operator $\mathcal{L}$ is an approximation of a projection whose kernel is given by the linear span of $\mathbf{L}$. $\mathds{1}-\mathcal{L}$, with $\mathds{1}$ being an identity matrix, is a projection onto the lipid subspace span($\mathbf{L}$) (see references for further details). A separate lipid projection operator is calculated for each subject. The network inputs are defined as
 
\begin{align}
    x_1 = m + l + w \\
    x_2 = (\mathds{1}-\mathcal{L}) x_1
\end{align}

The decontaminated solution ($\tilde{m}$) is obtained in two steps, first the network $\mathcal{Y}$ predicts the spectrum $y$, thereby approximating lipid $l$ and water $w$ signal. Subsequently, the spectrum $y$ is subtracted from $x_1$, the original water and lipid contaminated MRSI spectrum,
\begin{align}
    \mathcal{Y}(x_1,x_2) =& y \approx l + w \\
    \tilde{m} =& x_1 - y
\end{align}

A similar strategy is employed for the lipid removal only, with the exception that water signal is omitted.

\paragraph{Network Architecture:}

\begin{figure*}[t]
    \centering
    \includegraphics[width=12cm]{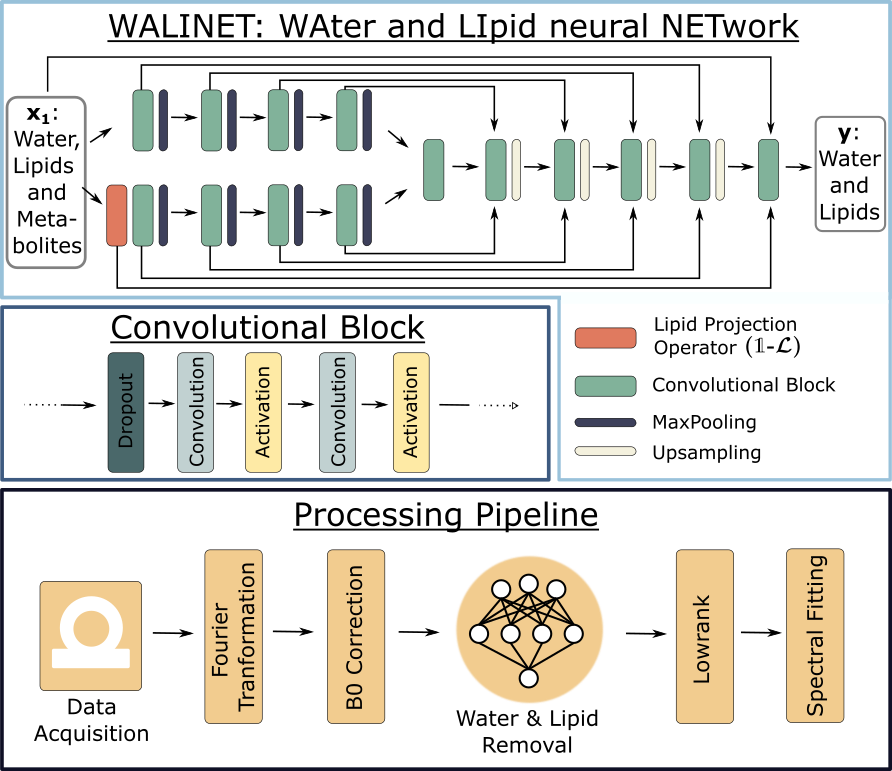}
    \caption{WAter and LIpid neural NETwork (WALINET): Top:  Y-Net architecture containing 4 convolutional blocks in each encoder and decoder, followed by a MaxPooling or Upsampling layer. Additional convolutional blocks are incorporated in the bottleneck and as a final layer. A lipid projection operator at the beginning of the second encoder enhances the distinguishability between metabolites and lipids. Bottom: The  WALINET is embedded into the MRSI processing pipeline shown at the bottom, which includes Fourier transformation, B0 correction, low-rank model and spectral quantification.}
    \label{fig:overview}
\end{figure*}

WALINET employs a Y-Net convolutional neural network structure \cite{mohammed2018net}, depicted in Figure \ref{fig:overview}, and characterized by the use of two encoders instead of one. Note, that the same network architecture can be used and trained only for lipid removal, which we call LIPNET (LIpid neural NETwork).

The Y-Net architecture enables enhanced contextual understanding by integrating features from different branches, leading to improved understanding of complex structures within the data. The results show an improvement compared to the U-Net \cite{mohammed2018net} and enable the encoding of different features in each encoder \cite{lan2020net}.

Each encoder ($\mathcal{E}$) and decoder ($\mathcal{D}$) of the Y-Net $(\mathcal{Y}$) comprised four convolutional blocks, each consisting of two convolutional layers, PReLU activation functions \cite{he2015delving}, dropout with a rate of 0.01, and MaxPooling/upsampling with a factor of 2. Skip connections are implemented between the encoders and decoder. An additional convolutional block is incorporated in the bottleneck region and after the decoder. The outputs of the encoders are concatenated and forwarded to the bottleneck convolutional block in the decoder.
\begin{align}
    \mathcal{Y}(x_1,x_2) =& \mathcal{D}(\mathcal{E}_1(x_1), \mathcal{E}_2(x_2)) \\
    =& \mathcal{D}(\mathcal{E}_1(x_1), \mathcal{E}_2((\mathds{1}-\mathcal{L}) x_1))
\end{align}

The kernel size of each block is set to 7. The number of channels in the first convolutional block is set to 16 and is doubled/halved after every MaxPooling/Upsampling layer.

\paragraph{Training Data:}
The training spectra were generated through a multi-step process, where metabolite spectra were simulated and combined with experimentally measured lipid and water spectra from \textit{in-vivo} human data. This approach was chosen because: 1) metabolite spectra can be realistically simulated for an extremely large range of very diverse parameters, 2) the lipid and water spectra are affected by a complex combination of experimental factors which are hard to be fully accounted in simulations, hence more realistic spectra can be extracted from measured data.  

Spectra for 25 common ${}^1$H metabolites were simulated using a physical model for the coupled spin systems\cite{Smith1994,songeon2023vivo}. An extensive dataset of $1.9\times10^6$ metabolite spectra was simulated for a wide range of concentrations, linewidths, noise levels, and baselines. Water and lipid signals were extracted from 19 subjects, including 2 glioma patients. The MRSI data were acquired with the 3D ${}^1$H-FID-ECCENTRIC sequence described in paragraph \ref{sec:data}). $10^5$ lipid \& water spectra were extracted from each subject, resulting in a total $1.9\times 10^6$ lipid \& water spectra that were randomly combined with metabolite spectra for training. Additionally, a validation dataset included 4 other subjects. Further details regarding generation of the training data is provided in Supplementary Information.

\paragraph{Training Procedure:}
The training spectra were subjected to further augmentation and normalization before being forwarded to the neural network. Online data augmentation during training was achieved by multiplication with a random phase 
\begin{align}
    \phi=e^{i\omega},~ \omega \in [0, 2\pi]
\end{align} 
to each spectrum. Normalization was performed dividing the input and ground truth spectra by an approximation of the energy $\mathds{E}$ of the underlying metabolic signal. Therefore, the root-sum-squared-error between the two input spectra was computed.
\begin{align}
    \mathds{E} = \sqrt{ \left| x_1 - x_2 \right|^T \left| x_1 - x_2 \right|}
\end{align} 

After augmentation and normalization, the spectra were separated into real and imaginary part, which were treated as separate channels during training. Mean-squared-error was employed as training loss, and computed on the separated real and imaginary channels.
\begin{align}
    \text{Loss} = \text{MSE}(\mathcal{Y}(\frac{x_1 \phi}{E}, \frac{x_2 \phi}{E}), \frac{y \phi}{E})
\end{align}

The network was trained for 400 epochs, using the Adam \cite{kingma2014adam} optimizer with a learning rate of 0.01, which was quartered every 50 epochs. The exponential decay rates of the first and second momentum of Adam $\beta_1$ \& $\beta_2$ were set to 0.9 and 0.999. 
The network was implemented using PyTorch 2.0.1 and CUDA 11.7 packages in Python 3.8.13. The model training was performed on a PowerEdge R7525 server (Dell) with 64 CPU cores (AMD EPYC7542 2.90GHz, 128M Cache, DDR4-3200), 512 GB CPU RAM (RDIMM, 3200MT/s), 3 GPU NVIDIA Ampere A40 (PCIe, 48GB GPU RAM) running Rocky Linux release
8.8 (Green Obsidian). 

\paragraph{Data Acquisition:} \label{sec:data}
\textit{In-vivo} MRSI data were acquired with 2D ${}^1$H-FID Cartesian phase encoded \cite{Klauser2021} and 3D ${}^1$H-FID ECCENTRIC \cite{klauser2023eccentric} pulse sequences using a 7T MR scanner (MAGNETOM Terra, Siemens Healthineers, Forchheim, Germany) and a 1Tx/32Rx head coil (NovaMedical, Wilmington, MA, USA).

2D ${}^1$H-FID Cartesian MRSI data were acquired on 2 subjects with matrix 53x41, 164x212 mm$^2$ field-of-view (FoV), 4x4 mm$^2$ in-plane voxel size, 10 mm slice thickness, spectral bandwidth of 4kHz, and 512 FID points. 

3D ${}^1$H-FID-ECCENTRIC \cite{klauser2023eccentric} was acquired on 21 subjects with 64x64x31 matrix, 220x220x105 mm$^3$ FoV, 3.4x3.4x3.4 mm$^3$ voxel size, spectral bandwidth of 2326 Hz, and 453 FID points.

For both sequences 0.9 ms echo-time (TE) and 275 ms repetition-time (TR) were used, resulting in 18min:40s for 3D ECCENTRIC and 7min:48s for 2D Cartesian. Further details are provided in Supplementary Material.

\paragraph{Processing Pipeline:}
The reconstruction and processing of 2D and 3D data is performed by a similar pipeline (Figure \ref{fig:overview}). However, the ECCENTRIC MRSI requires additional steps because of the non-Cartesian k-space sampling as further explained in the Supplementary Material. 

For comparison of WALINET and LIPNET performance, conventional nuisance signal removal was performed with HLSVD for water \cite{barkhuijsen1987improved} and L2 regularization for lipids. \cite{bilgic2014fast}.  The HLSVD retained the 32 largest eigenvalues of the Hankel matrix and the water removal was applied in the frequency range of 4.7ppm$\pm$0.5ppm. L2 regularization used lipid signals extracted from the skull mask of the subject with the regularization parameter $\beta$ individually adjusted for each subject to achieve a mean absolute diagonal value of its lipid suppression operator $\mathcal{L} = (\mathds{1}+\beta\mathbf{LL^H})^{-1}$ at an arbitrary value of 0.938, 
\begin{align}
\text{mean}(|\text{diag}(\mathcal{L})|) \sim 0.938.
\end{align}
This value was selected as the optimal trade-off between minimizing metabolite alteration and maximizing lipid suppression.

Following the water and lipid removal, a low-rank model was employed assuming separable spatial $U_n(r)$ and temporal $V_n(t)$ components of the metabolite signal,
\begin{align}
m(r,t)= \sum_{n=1}^{K} U_n(r)V_n(t)
\end{align}
with $K$ the rank of the model set to 40.
 
As final step, metabolic quantification was carried out by LCModel \cite{provencher1993estimation} spectral fitting.

%%==================================%%
%% Results %%
%%==================================%%

\section{Results}\label{sec2}

\begin{figure*}[t]
    \centering
    \includegraphics[width=.8\textwidth]{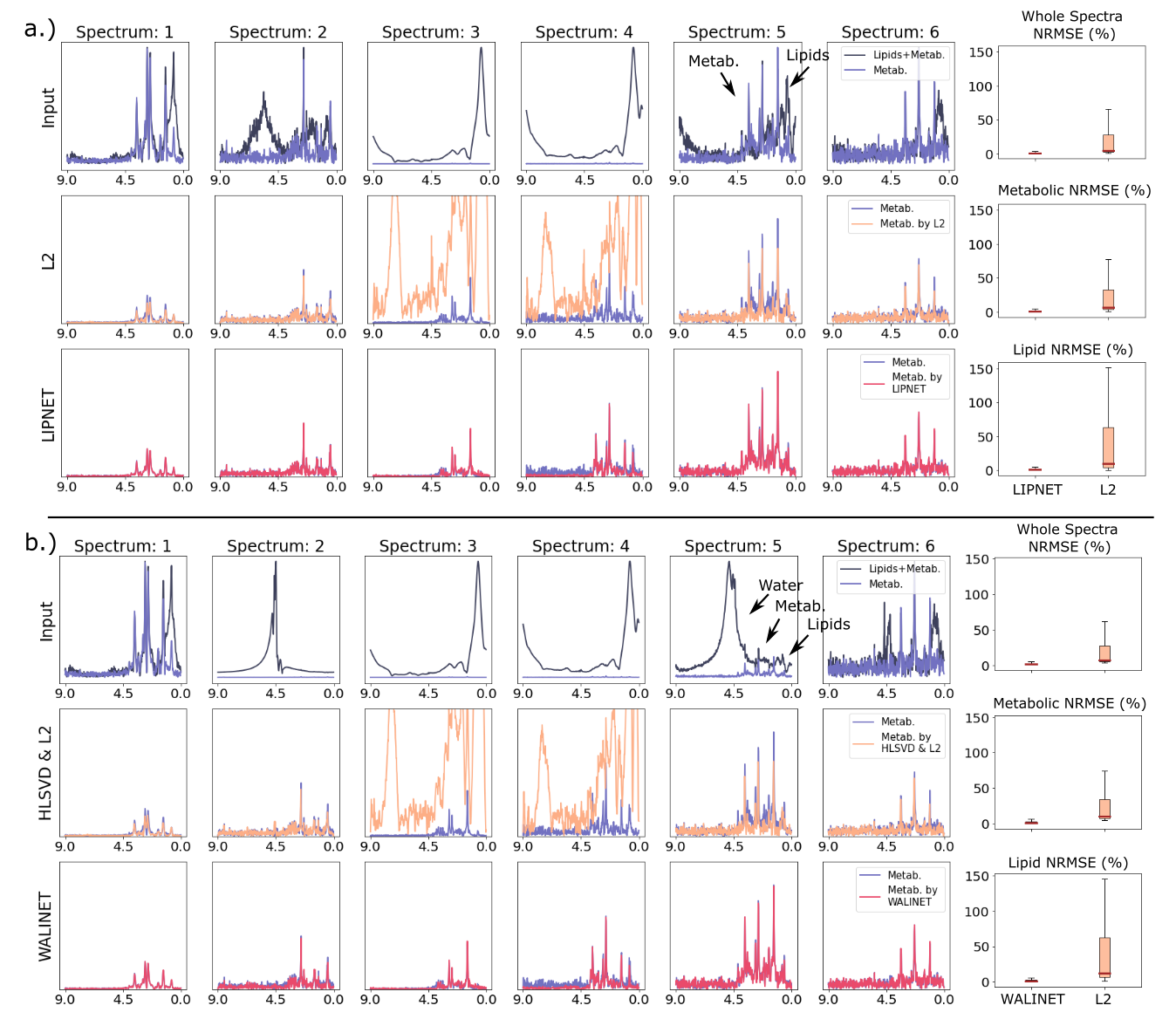}
    \caption{Simulation results. a) Comparison of lipid suppression by LIPNET and L2. Input spectra contaminated by lipids are shown in the first row (black), second row shows metabolite spectra recovered by L2 regularization (orange), \rev{third row shows the metabolite spectra recovered by LIPNET (red)}. b) Input spectra contaminated by water and lipids are shown on the first row (black), \rev{in the second row metabolite spectra processed with HLSVD+L2 are displayed,} and the metabolite spectra predicted by WALINET (red) are plotted below. The ground-truth metabolite spectra (blue) are overlaid in all spectral plots. Normalized root mean squared error (NRMSE) is computed for the whole spectrum \rev{(9.0-0.0 ppm)}, the metabolic range (4.2-1.9 ppm) and the lipid range (1.9-0.7 ppm) for each method LIPNET or L2 i\rev{n a) and WALINET or HLSVD+L2 in b)}. Separate evaluation of specific spectral ranges allows an individual assessment of the preservation of metabolic signals, as well as the effectiveness of lipid and water suppression. Arrows indicate the position of the main peaks of water, lipids and metabolites.}
    \label{fig:simulated}
\end{figure*}

\paragraph{Simulation Results}
In Figure \ref{fig:simulated} the results of LIPNET and WALINET on simulated data are compared to L2 and HLSVD+L2, respectively. Evaluation data were created by merging 100,000 simulated metabolite spectra with \textit{in-vivo} water \& lipid  signals (for WALINET) or only lipid signals (for LIPNET) extracted from a subject excluded from the training data. Results on evaluation spectra contaminated only by lipid are shown in Figure \ref{fig:simulated}a, and by water \& lipid are shown in Figure \ref{fig:simulated}b. We obtained more agreement between ground truth metabolite spectra and the predicted metabolite spectra in the case of LIPNET and WALINET than in the case of L2 and HLSVD+L2 regularization. Spectra obtained by L2 regularization tend to show more residual lipid signal and more suppression of the NAA peak compared to WALINET and LIPNET. Quantitatively, boxplots of the normalized root mean square error (NRMSE) show that WALINET and LIPNET remove more lipid signal (interquartile NRMSE 0.86\text{\%}-2.69\text{\%}) while preserving more metabolite signal (interquartile NRMSE 0.62\text{\%}-1.45\text{\%}) compared to L2 (lipid interquartile NRMSE 3.68\text{\%}-6.45\text{\%} and metabolite interquartile NRMSE 1.04\text{\%}-4.11\text{\%}).

%Any lipid suppression performed by a linear operator, such as the L2-lipid-regularization, is limited in its complexity, hence requires a trade-of in suppression strength. Therefore, lipid suppression of L2-lipid-regularization requires a balance between minimizing the suppression of metabolic signal and the maximization of the removal of remaining lipid signal. 

%+++ Full Range +++
%MEAN
%WALINET:  2.5902526
%LIPNET:  1.6317365
%L2:  6.7512603
%MEDIAN
%WALINET:  1.9573159
%LIPNET:  1.1692096
%L2:  3.4538693
%+++ Metabolic Range +++
%MEAN
%WALINET:  2.985007
%LIPNET:  2.101612
%L2:  7.6211195
%MEDIAN
%WALINET:  2.2264488
%LIPNET:  1.3961518
%L2:  4.896678
%+++ Lipid Range +++
%MEAN
%WALINET:  1.6671633
%LIPNET:  1.1377628
%L2:  5.492407
%MEDIAN
%WALINET:  1.249726
%LIPNET:  0.9271945
%L2:  1.5923471

\paragraph{\textit{In-vivo} Results}
\textit{In-vivo} performance was tested on Cartesian encoded 2D MRSI and ECCENTRIC encoded 3D MRSI data from human volunteers. The speed of all methods for nuisance signal removal was timed on \textit{in-vivo} high-resolution 3D MRSI data and showed considerable faster times for WALINET compared to HLSVD+L2, as listed in Table \ref{tab:times}. The 2D MRSI was used to test the generalizability of WALINET and LIPNET trained on 3D MRSI data.

\begin{table*}[t]
\begin{center}
%\begin{wraptable}{t}{7cm}
\begin{tabular}{lcccc}\\\toprule  
 & \makecell{WALINET \\(mm:ss)} & \makecell{HLSVD+L2 \\ (mm:ss)} & \makecell{HLSVD \\ (mm:ss)} & \makecell{L2 \\ (mm:ss)}\\\midrule
10 cores & 00:08 & 03:08 & 03:05 & 00:03 \\  \midrule
1 core & 00:08 & 42:49 & 42:46 & 00:03 \\  \bottomrule
\end{tabular}
\caption{Processing times of WALINET, water HLSVD and lipid L2 regularization for 1 CPU core and parallelized for 10 CPU cores. LIPNET is based on the same Y-NET architecture as WALINET and therefore requires the same processing time. Parallelization of L2 and WALINET across several cores is not possible, therefore equivalent processing times are given in each row for these two methods.}
\label{tab:times}
\end{center}
\end{table*}
%\end{wraptable} 

\begin{figure*}[ht]
    \centering
    \includegraphics[width=1\textwidth]{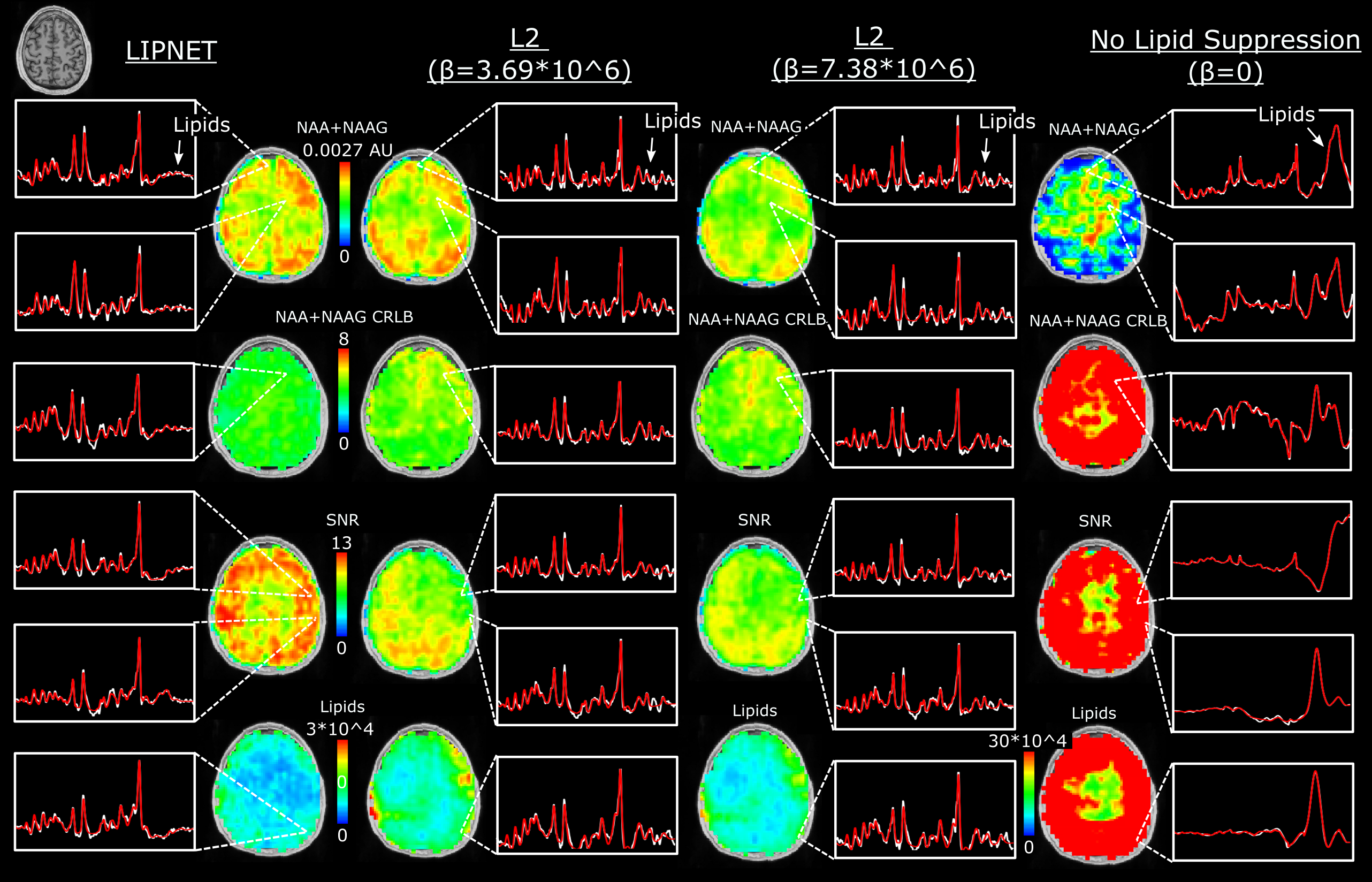}
    \caption{Comparison of lipid removal on \textit{in-vivo} 2D MRSI by LIPNET and L2 for three values of the regularization parameter $\beta$, the optimal value ($3.69*10^6$), double the optimal value, and zero for no lipid suppression. Metabolic maps are shown for NAA+NAAG, the corresponding CLRB, SNR computed by LCModel \cite{provencher1993estimation} and residual lipid signal. Spectra from several brain voxels are shown for each method, the white trace shows the measured spectrum, the red trace shows LCModel fit.}
    \label{fig:qualiLip}
\end{figure*}

First, we studied the effects of different lipid suppression methods. Figure \ref{fig:qualiLip} compares lipid removal by LIPNET and L2 regularization on \textit{in-vivo} 2D MRSI data, while the water removal has been done by HLSVD for all the tests. For L2 two regularization parameters were used: 1) $\beta=3.69*10^6$ optimized for the 0.938 mean absolute diagonal value of the lipid suppression operator, and 2) $\beta=7.38*10^6$ which doubles lipid regularization parameter for stronger lipid suppression. In addition, results obtained with no lipid suppression ($\beta=0$) are presented. 

Metabolic maps of NAA+NAAG obtained with lipid suppression methods show similar structural features with good contrast between gray-white matter. However, some differences can be observed: 1) L2 regularization produces lower levels for NAA+NAAG maps compared to NAA+NAAG levels obtained by LIPNET, 2) the metabolic maps obtained with the strongest L2 regularization ($\beta=7.38*10^6$) have the lowest metabolite levels. The superior performance of LIPNET is also confirmed by the spectral quality maps, which show 10\text{\%}-50\text{\%} smaller CRLB and 155\text{\%} higher SNR compared to L2 regularization. Lipid maps show 60\text{\%} lower residual lipid signal for LIPNET compared to L2 regularization. On the other hand, it can be seen that without any lipid removal the metabolic maps are completely overwhelmed by lipid artifacts with no visible structural details of the brain and very large quantification errors. The lipid  Examples of spectra show clearly more residual lipid signal by L2 regularization than by LIPNET, while in the case of no lipid removal the metabolite spectra are heavily distorted by the large lipid signal. In addition, metabolic maps of Cr+PCr, and Glutamate, which are consistent with the previous findings, are presented in Supplementary Figure \ref{fig:qualiLipv2}.

\begin{figure*}[t]
    \centering
    \includegraphics[width=1\textwidth]{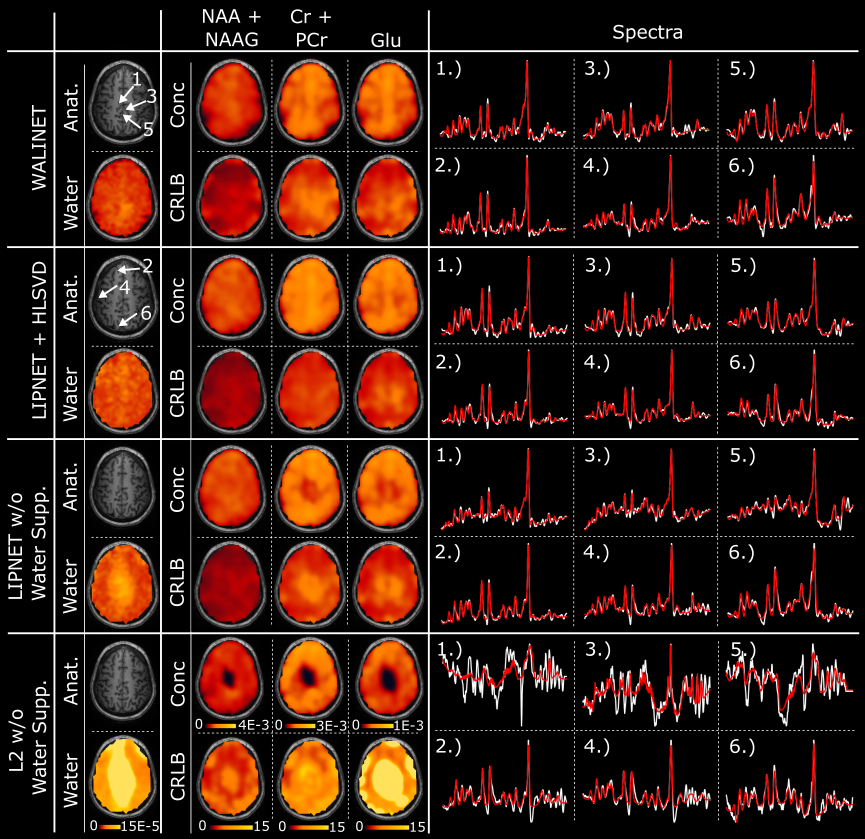}
    \caption{Comparison of water removal methods on \textit{in-vivo} 2D MRSI, including WALINET, HLSVD+LIPNET, LIPNET and L2 without water suppression. WET\cite{ogg1994wet} water suppression  was used during acquisition. Metabolic maps and the corresponding CRLBs are shown for NAA+NAAG, Cr+PCr and Glu together with maps of the residual water signal and examples of spectra (voxel locations indicated by arrows, red line indicates LCModel \cite{provencher1993estimation} fit and the white line the experimental spectra).}
    \label{fig:qualiWali}
\end{figure*}

Second, we studied the effects of different water removal methods. Figure \ref{fig:qualiWali} compares the water removal by WALINET and HLSVD on \textit{in-vivo} 2D MRSI. Similar maps of the residual water signal and metabolites are obtained for WALINET and LIPNET+HLSVD. The maps obtained without water suppression show higher residual water signal and signal dropout in the center of the brain, which is worse for L2 than LIPNET.  Spectra and LCModel \cite{provencher1993estimation} fit show progressively larger baseline noise and more distortion of metabolite peaks when going from WALINET to LIPNET+HLSVD, LIPNET, and L2. \rev{Examples of spectra with full spectral window and intensity range are shown in Supplementary Figure \ref{fig:FRspectra}.} 

Taken in combination, the results from Figures \ref{fig:qualiLip} and \ref{fig:qualiWali} indicate that WALINET and LIPNET generalize well to different acquisition schemes (2D Cartesian vs. 3D Non-Cartesian) that were not used for the acquisition of training data.

\begin{figure*}[t]
    \centering
    \includegraphics[width=1\textwidth]{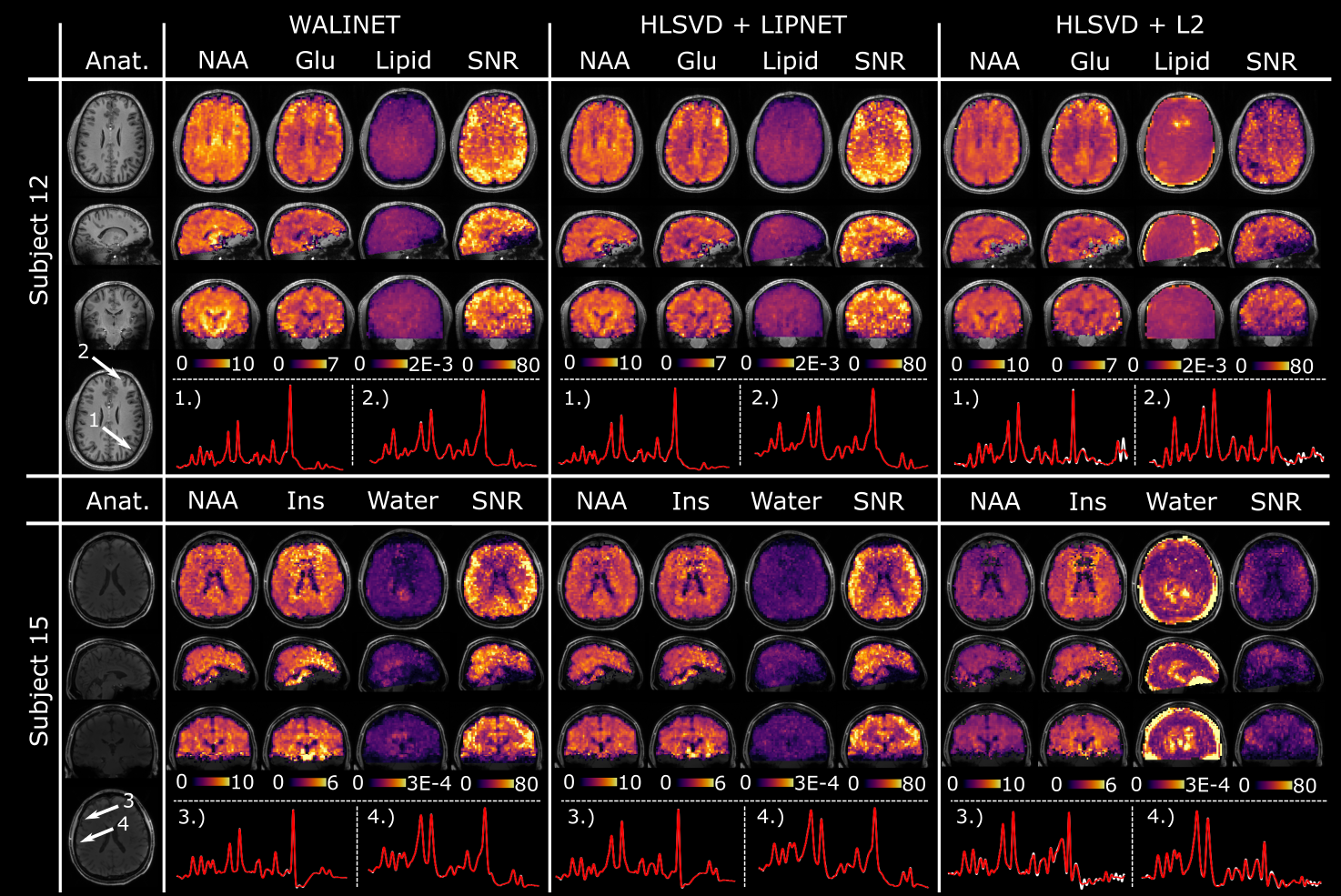}
    \caption{Comparison of combined water \& lipid removal on \textit{in-vivo} 3D MRSI. Results are shown for WALINET, HLSVD+LIPNET and HLSVD+L2, including metabolic maps for NAA, Glutamate, Inositol, residual lipid signal, residual water signal, and SNR computed by LCModel. Selected spectra from individual voxels indicated by white arrows on the anatomical images are shown at the bottom (white trace shows measured spectrum, red trace shows LCModel \cite{provencher1993estimation} fit).}
    \label{fig:qualComp3D}
\end{figure*}

Third, we investigated the effects of combined water and lipid removal on 3D MRSI. Figure \ref{fig:qualComp3D} compares the performance of WALINET, HLSVD+LIPNET and HLSVD+L2 on \textit{in-vivo} 3D MRSI data from two evaluation subjects. It can be seen that WALINET and HLSVD+LIPNET provide similar metabolic maps, residual lipid \& water maps, SNR and spectra. The results provided by HLSVD+L2 show more residual lipid \& water signal, lower SNR, lower gray/white matter contrast in metabolite maps. In particular, spectra obtained by  HLSVD+L2 show a reduction of the NAA peak and larger residual lipid peaks.

%\paragraph{A quantitative study evaluating the spectral quality of presented methods}

\begin{figure*}[t]
    \centering
    \includegraphics[width=1\textwidth]{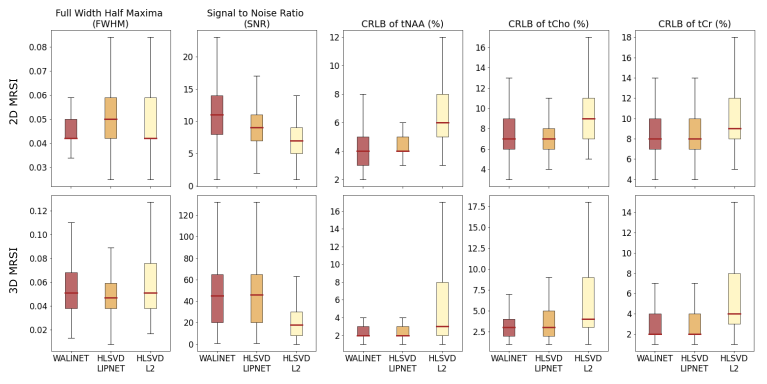}
    \caption{Boxplots of spectral quality metrics obtained be WALINET, HLSVD+LIPNET and HLSVD+L2 in 2D and 3D \textit{in-vivo} MRSI data.}
    \label{fig:boxplot}
\end{figure*}

Figure \ref{fig:boxplot} compares the metrics of spectral quality obtained by WALINET, HLSVD+LIPNET, and L2+HLSVD on 2D and 3D datasets. Results indicate that WALINET has similar mean SNR (11/45 in 2D/3D) and mean CRLB (NAA/Cho/Cr = 4/7/8\text{\%} in 2D and 2/3/2\text{\%} in 3D) compared to HLSVD+LIPNET (SNR = 9/46 in 2D/3D; CRLB of NAA/Cho/Cr = 4/7/8\text{\%} in 2D and 2/3/2\text{\%} in 3D). At the same time, both of these deep learning-based methods have higher SNR and lower CRLB compared to conventional L2+HLSVD (SNR = 7/18 in 2D/3D; CRLB of NAA/Cho/Cr = 6/9/9\text{\%} in 2D and 3/4/4\text{\%} in 3D). Additionally, the interquartile interval and the min-max whiskers of CRLB are narrower for WALINET and HLSVD+LIPNET than HLSVD+L2. The spectral linewidths show similar mean values (0.04-0.05 ppm) for all three methods, but narrower interquartile interval and min-max whiskers for WALINET and HLSVD+LIPNET than HLSVD+L2. 

\rev{The lipid removal factor and water removal factor was quantified \textit{in-vivo} and simulations for each of the methods and listed in Supplementary Table \ref{tab:lipTable}. It can be seen that WALINET and LIPNET provide \textit{in-vivo} a mean lipid removal factor between 25-45 in the brain (689-876 lipid removal factor in the scalp), and a mean water removal factor between 34-53.} 

%2D
%+++ WALINET +++
%snr:  11.0
%crlb-naa:  4.0
%crlb-cho:  7.0
%crlb-cr:  8.0
%+++ LIPNET +++
%snr:  9.0
%crlb-naa:  4.0
%crlb-cho:  7.0
%crlb-cr:  8.0
%+++ L2 +++
%snr:  7.0
%crlb-naa:  6.0
%crlb-cho:  9.0
%crlb-cr:  9.0

%3D
%+++ WALINET +++
%snr:  45.0
%crlb-naa:  2.0
%crlb-cho:  3.0
%crlb-cr:  2.0
%+++ LIPNET +++
%snr:  46.0
%crlb-naa:  2.0
%crlb-cho:  3.0
%crlb-cr:  2.0
%+++ L2 +++
%snr:  18.0
%crlb-naa:  3.0
%crlb-cho:  4.0
%crlb-cr:  4.0

%%==================================%%
%% Discussion %%
%%==================================%%

\section{Discussion}\label{sec3}

We demonstrate WALINET, a fast and robust nuisance signal identification convolutional neural network. WALINET was trained to identify water and lipid signals in whole-brain ${}^1$H-MRSI spectra, and simultaneously removes these signals to allow accurate quantification of metabolites. WALINET eliminates time-consuming computations for water removal and iterative single-subject hyperparameter optimization for lipid suppression by conventional methods.  Thereby, WALINET can streamline and automate MRSI data processing for user-friendly clinical applications. 

Considering that the evolution of MRSI is towards high-spatial resolution with large matrix size, the development of computationally efficient processing pipelines is required to keep up with the computational demands posed by the need to process increasing data size. Our evaluation showed that WALINET is considerably faster (8s) on high-resolution MRSI compared to conventional methods (42min). 

In addition to faster processing times, WALINET showed superior performance with more lipid removal and preserving more metabolite signal compared to state-of-the-art conventional lipid removal methods. WALINET provided significantly higher data quality, effectively doubling the SNR and lowering by half quantification errors of metabolites, compared to conventional methods.  \rev{WALINET and LIPNET have a lipid removal factor that is 11.19 times larger than L2, and a water removal factor that is 2.86 times larger compared to HLSVD. Compared to other methods listed in the recent consensus paper \cite{tkavc2021water}, WALINET and LIPNET provide lipid suppression factors similar to the crusher coil and ECLIPSE but without the need of special hardware. A potential caveat of WALINET and LIPINET is the lactate doublet at 1.3 ppm may be removed together with the lipid signal, hence the ability of to obtain lactate maps is dependent on the lactate quadruplet at 4.1 ppm. By comparison the nuisance signal removal using the union-of-space model \cite{ma2016removal} has been shown to retain the lactate peak at 1.3 ppm.} 

Convolutional neural networks act as nonlinear functions that may model better lipid contamination in MRSI, while conventional methods \cite{bilgic2014fast,ma2016removal} that assume a linear orthogonal relationship between lipids and metabolites may inadvertently  remove metabolite signal when this assumption is not met. The effects of improved metabolite quantification translate into metabolite images that have better structural details. We believe that a robust self-contained efficient nuisance signal removal method that is implemented as an independent processing step is very useful and can be combined with any MRSI processing pipeline. This may offer greater flexibility compared to methods that are fully embedded with reconstruction of k-space MRSI data. 

While LIPNET and WALINET were exclusively trained on 3D MRSI spectra, the presented results demonstrate robust performance also on 2D MRSI test data, which differs in FID length and spectral bandwidth (echo spacing). In our experience, machine learning algorithms are able to extrapolate to a certain extent to out-of-distribution data. However, the extension of WALINET to different sequences, spectral bandwidths and FID lengths remains future work.

At the moment, the demonstration of WALINET performance was limited to 7T ultra-high-field MRSI, which is the highest field approved for clinical use. Methods such as WALINET are highly relevant at 7T because the high SAR and non-uniform $B_0$ and $B_1+$ fields make pulse sequence-based suppression of water and lipids highly impractical for whole-brain MRSI. However, we expect that the approach and same network can be employed with additional training at lower (3T) or higher fields. We also expect that the processing time of WALINET will be similar for higher spatial resolution of MRSI, while the time for conventional methods will linearly scale with the data size. Furthermore, removal of lipid signal is extremely important for accelerated undersampled MRSI acquisitions \cite{nassirpour2018multinet,nassirpour2018compressed} where aliased lipid signal can overwhelm parallel imaging and compressed sense reconstructions. Based on its performance, WALINET has great potential for being used in combination with undersampled MRSI, and we will explore this in future work. This is suggested by the generalization of WALINET from 3D Non-Cartesian sampling to 2D Cartesian sampling that we demonstrated in this paper. In addition, we showed that the same model can be used for lipid-only removal (LIPNET), which can be further combined with other processing methods. We also noticed that LIPNET provides some suppression of the water tail in the metabolite spectral range. This happens because there are lipid signals (at 5.1ppm and 4.4 ppm) close to the water peak which the network learns to remove. However, in the presence of a large residual water peak LIPNET is not sufficient and a water trained network is necessary such as WALINET.

\paragraph{Conclusion}
Efficient removal of water and lipid signals is key for proton MRSI-based quantitative metabolic imaging. Convolutional neural networks such as WALINET provide an effective approach towards this goal with superior performance compared to conventional methods. We provide WALINET, including the computation of the L2 lipid operator, as a self-contained package that can be used as a plug-in with any MRSI processing pipeline. Since WALINET does not need specialized hardware, it can be easily disseminated to automate the MRSI processing pipeline for high-throughput workflow in clinical applications. We anticipate that these aspects will lead to larger adoption and impact of MRSI for clinical applications and research.
\newline

\textbf{Data Availability Statement.} Testing data can be obtained from the authors based on reasonable request and institutional approved data sharing agreement. The code for WALINET is publicly available at www.github.com/weiserjpaul/WALINET. 

\section*{Acknowledgments}

This work was supported by National Institute of Health research grants 2R01CA211080-06A1, R01CA255479, P50CA165962, P41EB015896, R00HD101553, the Athinoula A.\ Martinos Center for Biomedical Imaging,  Austrian Science Fund: WEAVE I 6037 \& P 34198, and the Christian Doppler Laboratory for MR Imaging Biomarkers (BIOMAK).

\clearpage
\newpage

\bibliography{sections/MRM-AMA}
\newpage

%\section*{Supplemetary Material}
%\centering{\Huge{\textbf{Supplementary Material}}}
\twocolumn[\section*{\centering{\Huge{\textbf{Supplementary Material}}}}]
\renewcommand{\figurename}{\textbf{Supporting Figure}}
\renewcommand{\thefigure}{S\arabic{figure} }
\setcounter{figure}{0}

\renewcommand{\tablename}{\textbf{Supporting Table}}
\renewcommand{\thetable}{S\arabic{table} }
\setcounter{table}{0}

\section{Supplementary Methods}

\paragraph{Data Acquisition:}
2D 1H-FID Cartesian MRSI data were acquired with unaccelerated elliptical phase-encoded. 3D 1H-FID-ECCENTRIC \cite{klauser2023eccentric} was acquired  with randomly positioned circular trajectories with the radius set to kmax/8 without temporal interleaving and full sampling (AF=1) of a spherical 3D k-space. 

For both sequences the excitation was performed with a Shinnar-LeRoux optimized pulse \cite{Klauser2021, klauser2023eccentric} having 6.5kHz bandwidth, 1ms duration and 27° excitation flip-angle. No lipid suppression was employed in the sequence, while water suppression was achieved by a four-pulses WET method \cite{Klauser2021, klauser2023eccentric}.

In addition, low-resolution water-unsuppressed MRSI were acquired as calibration scan using the same sequences but omitting WET and with a smaller matrix (22x22x11 for 3D and 22x22 for 2D) in 1:16 min:s for 3D ECCENTRIC and 2:12 min:s for 2D Cartesian. The water-unsuppressed MRSI were used for coil combination and $B_0$ field inhomogeneity correction.

\paragraph{Training Data:}
First, spectra for 25 common ${}^1$H metabolites were simulated using a physical model for the coupled spin systems\cite{Smith1994,songeon2023vivo}. An extensive dataset of $1.9\times10^6$ metabolite spectra was simulated for a wide range of concentrations, linewidths, noise levels, and baselines. 
The metabolite concentrations were distributed according to a normal distribution with a standard deviation five times greater than the mean of 1 (arbitrary units), with truncation to zero to ensure positive values. The simulation included also variations in frequency offset (-150 to 150 Hz), Voigt linewidth (4 to 50 Hz), and signal-to-noise ratio (SNR from 1 to 10). Random baselines were introduced, characterized by 10 broad gaussian components. 

Second, we extracted a collection of lipid spectra from voxels within the scalp region obtained from \textit{in-vivo} MRSI datasets. Representative head/brain/scalp masks and B0 field maps are presented in Supplementary Figure \ref{fig:mask}. The amplitude of lipid signal from the lipid mask was varied with a scaling factor, spanning a broad interval from $10^{-2}$ to $10^3$. \rev{In addition, the phase of the lipid signal was varied between $-\pi$ and $+\pi$. The variation of the amplitude and phase of the lipid signal mimics the sinc ringing of the point spread function. The large range of amplitude and phase variation effectively augments the \textit{in-vivo} lipid contamination, which makes the network robust to different k-space sampling strategies. Hence, we do not need to retrain the network when changing the acquisition. This was verified by testing WALINET on MRSI data acquired with ECCENTRIC as well as cartesian phase encoding.}

Third, the water signal was extracted using Hankel-Lanczos singular value decomposition (HLSVD) with a rank of 64 from voxels within the brain from the same \textit{in-vivo} MRSI datasets. To further augment the \textit{in-vivo} water distribution the 10 water signal components estimated by HLSVD in each voxel were randomly weighted with a factor ranging from $10^{-1}$ to $10^2$. 

To create the final training input spectra ($x_1$) for WALINET the simulated metabolite spectra ($m$) and experimentally-derived lipid \& water spectra ($x_2$) were randomly combined. Note, that in the case of LIPNET the input training spectra combined simulated metabolite and experimentally-derived lipid spectra, without the water signal. 

 Water and lipid signals were extracted from 19 subjects, including 2 glioma patients. The MRSI data were acquired with the 3D ${}^1$H-FID-ECCENTRIC sequence described in the Methods. $10^5$ lipid \& water spectra were extracted from each subject, resulting in $1.9\times 10^6$ total metabolite + lipid \& water spectra used for training. Additionally, a validation dataset included 4 other subjects.

\paragraph{Processing Pipeline:}
For 3D ECCENTRIC the k-space sampling density was compensated based on Voronoi diagrams \cite{malik2005gridding} where each k-space point is normalized by the area of its assigned Voronoi vertex. Upon the weighting of the k-space data, an inverse non-uniform discrete Fourier transform (iNUFT) was applied for each stack (kx-ky plane) within the k-space domain. Subsequently, an additional inverse fast Fourier transform (iFFT) was performed along the kz dimension to finalize the reconstruction of the MRSI data. In contrast, for 2D ${}^1H$-FID-MRSI data a iFFT is applied on the 2D k-space datasets. For both 2D and 3D MRSI a Hamming filter was applied in k-space prior to the Fourier transform. After the transformation from k space to image space, the spectra are obtained by Fourier transform of the time dimension. Coil combination was performed with ESPIRIT \cite{uecker2014espirit} using sensitivity profiles computed from water un-suppressed MRSI. A correction for $B_0$ field inhomogeneity (Supp. Fig. \ref{fig:mask}) is computed from water un-suppressed data and applied to the coil-combined image space MRSI data, which was followed by water and lipid removal. 

LCModel used a basis set simulated by NMR quantum mechanics in GAMMA \cite{smith1994computer} for twenty-two metabolites: phosphorylcholine (PCh), glycerophosphorylcholine (GPC), creatine (Cr), phosphocreatine (PCr), gamma-aminobutyric acid (GABA), glutamate (Glu), glutamine (Gln), glycine (Gly), glutathione (GSH), myo-inositol (Ins), N-acetylaspartate (NAA), N-acetyl aspartylglutamate (NAAG), scylloinositol (Sci), lactate (Lac), threonine (Thr), beta-glucose (bGlu), alanine (Ala), aspartate (Asp), ascorbate (Asc), serine (Ser), taurine (Tau), and 2-hydroxyglutarate (2HG) and a measured macromolecular background \cite{povavzan2015mapping}. Note that during training WALINET and LIPNET learn to remove macromolecule signal, since this is present together with lipid signal in the scalp spectra used to generate training data. This was verified experimentally, as macromolecular fitting by LCModel was very close to 0 throughout the brain. The spectral fitting was done for the 1ppm-4.2ppm spectral range and the results for each voxel were used to generate metabolic images. The unsupressed water reference signal was used as quantification reference for metabolites concentrations (institutional units, I.U.) to compare metabolite levels across subjects and scanners. To assess the quality of the MRSI data and fit, linewidth (FWHM), signal-to-noise ratio (SNR), and Cramer-Rao lower bounds (CRLB) goodness of fit maps were generated.

\rev{The lipid removal factor was calculated in test subjects and in simulations. For this we integrated the absolute value of the lipid signal before and after removal: a) in the range 0.8-1.8 ppm to exclude the metabolite region (1.8-4.2 ppm) for the in-vivo data, and b) over the entire spectral range of interest 0.8-4.2 ppm in simulations, since in simulations we have the metabolite-free lipid signal over the entire range. The water removal factor was calculated by integrating the absolute water signal between 4.2-5.2 ppm, before and after the removal.}

\begin{figure*}[t]
    \centering
    \includegraphics[width=.9\textwidth]{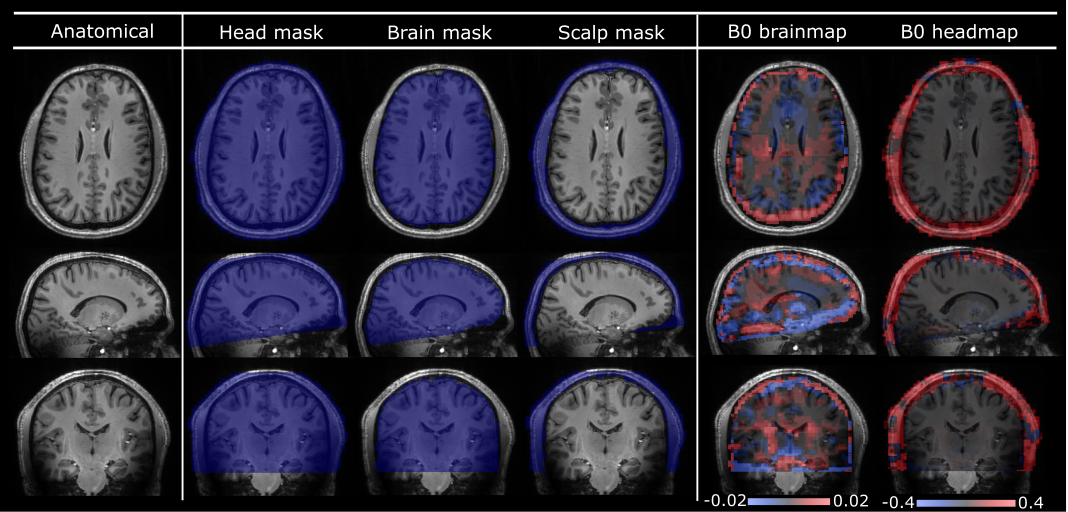}
    \caption{Examples of head/brain/scalp binary masks and the corresponding B0 field maps (specified in ppm).}
    \label{fig:mask}
\end{figure*}

\begin{figure*}[t]
    \centering
    \includegraphics[width=1\textwidth]{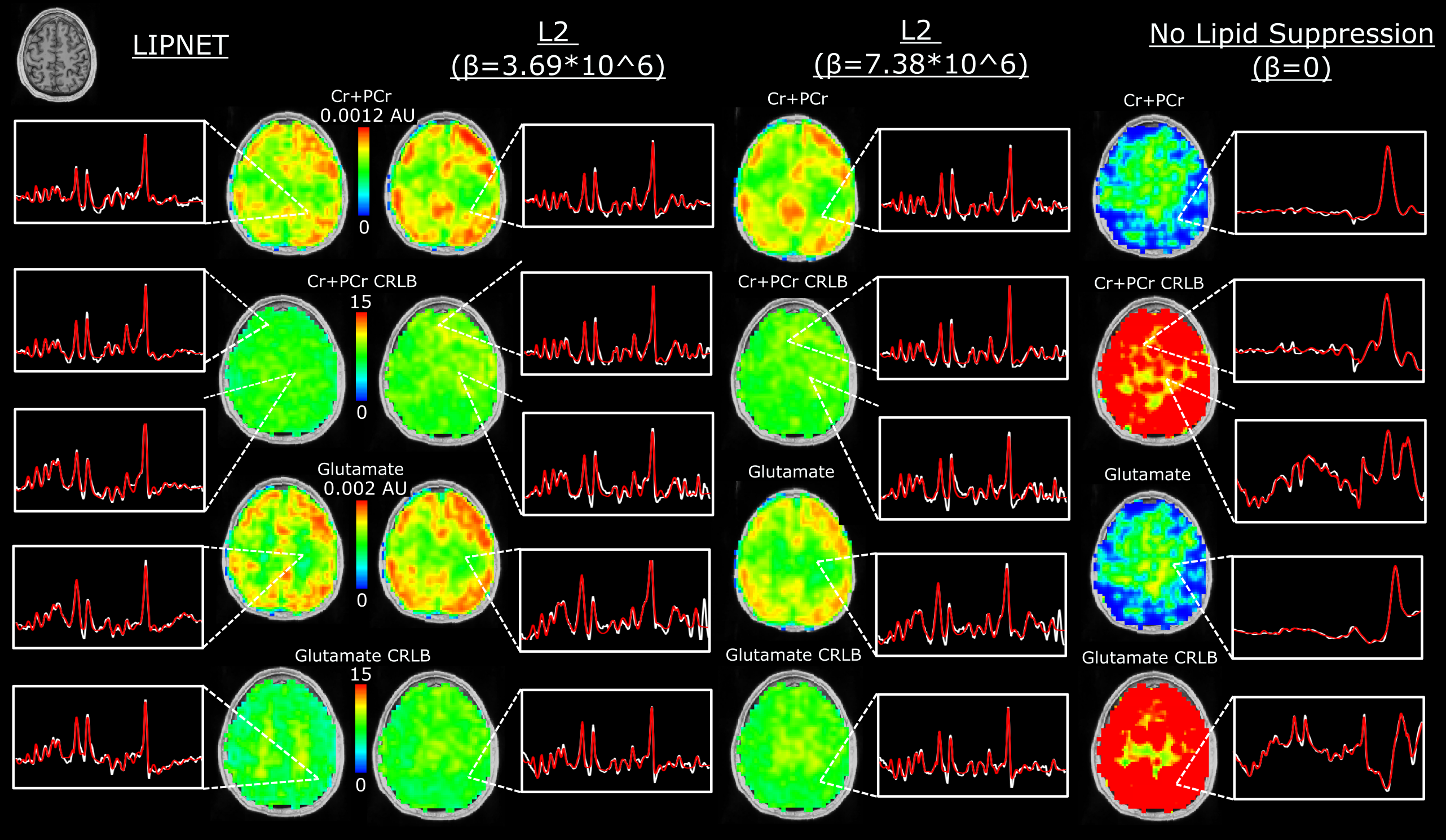}
    \caption{Comparison of lipid removal on \textit{in-vivo} 2D MRSI by LIPNET and L2 for three values of the regularization parameter $\beta$, the optimal value ($3.69*10^6$), double the optimal value, and zero for no lipid suppression. Maps are shown for two metabolites Cr+PCr, Glu and their corresponding CLRB. Spectra from several brain voxels are shown for each method, the white trace shows the measured spectrum, the red trace shows LCModel fit.}
    \label{fig:qualiLipv2}
\end{figure*}

\begin{figure*}[t]
    \centering
    \includegraphics[width=1\textwidth]{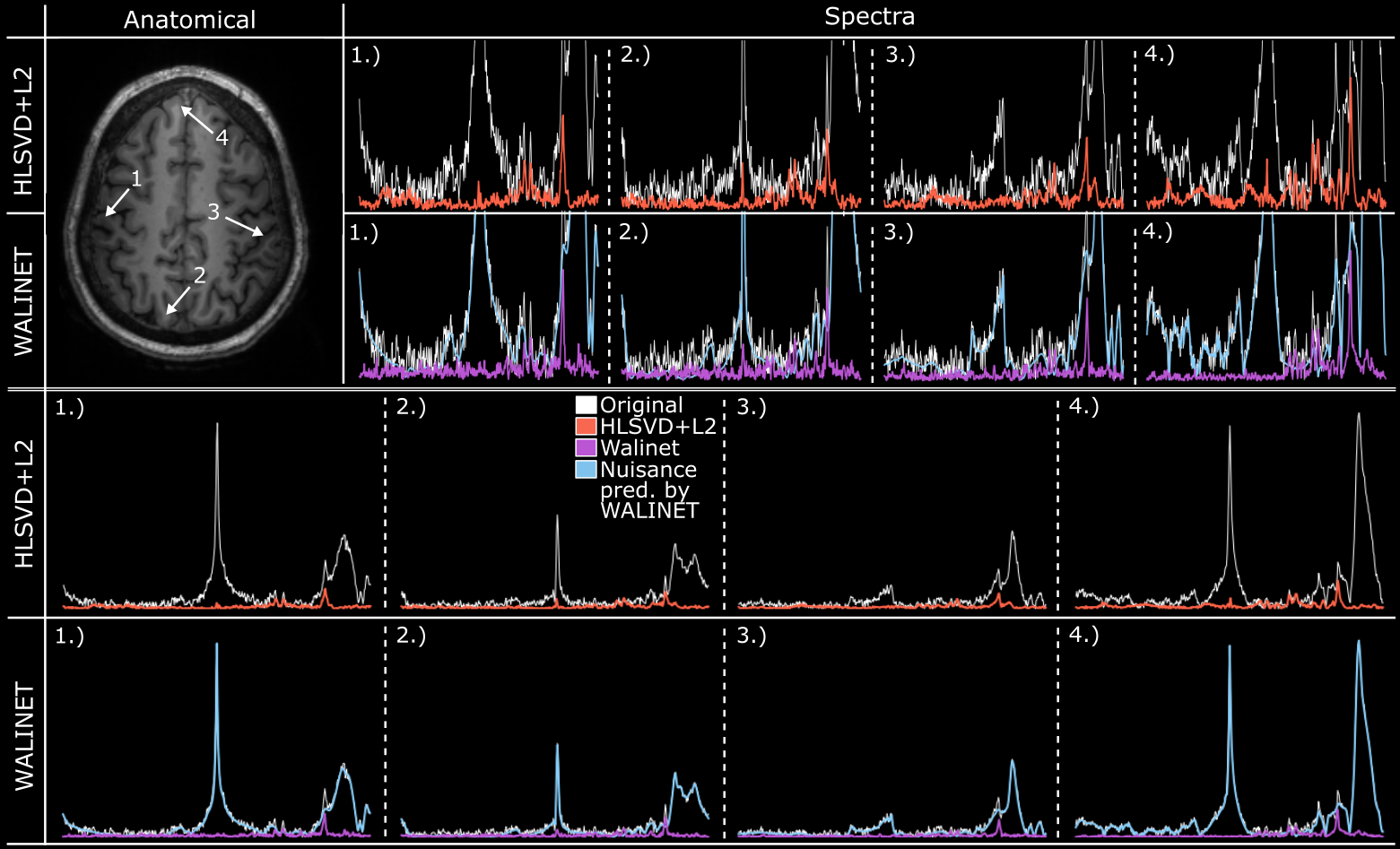}
    \caption{\rev{Comparison of water and lipid removal to WALINET on full range spectra. Each specrtum is shown enlarged on top and in full size below. The nuisance contaminated spectra is displayed in white, HLSVD+L2 processed spectra in orange, and WALINET nuisance removed spectra in purple. The from WALINET predicted nuisance signal is overlayed in light blue color.}}
    \label{fig:FRspectra}
\end{figure*}

\begin{table*}[!ht]
    \centering
    \begin{tabular}{|c|c|c|c|c|}
        \hline
                                                            &                       & vol12 & vol15 & Simulated \\ \cline{1-5}
        \multirow{4}{*}{\shortstack{\textbf{Lipid Removal Factor:} \\Brain mask} }  
                                                            & \multirow{2}{*}{Mean} & \cellcolor{customRed!80}25.43 
                                                            & \cellcolor{customRed!80}45.28 & \cellcolor{customRed!80}235.55    \\ \cline{3-5}
                                                            &                       & \cellcolor{customYellow!80}20.72 
                                                            & \cellcolor{customYellow!80}25.61 & \cellcolor{customYellow!80}17.68  \\ \cline{2-5}
                                                            &\multirow{2}{*}{Median}& \cellcolor{customRed!80}20.57 
                                                            & \cellcolor{customRed!80}35.60 & \cellcolor{customRed!80}118.51    \\ \cline{3-5}
                                                            &                       & \cellcolor{customYellow!80}18.07 
                                                            & \cellcolor{customYellow!80}22.41 & \cellcolor{customYellow!80}8.97  \\ \hline
        \multirow{4}{*}{\shortstack{\textbf{Lipid Removal Factor:} \\Head mask}}   
                                                            & \multirow{2}{*}{Mean} & \cellcolor{customRed!80}689.23 
                                                            & \cellcolor{customRed!80}876.14 & \cellcolor{customRed!80}235.55  \\ \cline{3-5}
                                                            &                       & \cellcolor{customYellow!80}76.81 
                                                            & \cellcolor{customYellow!80}73.59 & \cellcolor{customYellow!80}17.68 \\ \cline{2-5}
                                                            &\multirow{2}{*}{Median}& \cellcolor{customRed!80}78.69 
                                                            &\cellcolor{customRed!80}156.73 & \cellcolor{customRed!80}118.51   \\ \cline{3-5}
                                                            &                       & \cellcolor{customYellow!80}33.56 
                                                            & \cellcolor{customYellow!80}37.36 & \cellcolor{customYellow!80}8.97  \\ \hline
        \multirow{4}{*}{\shortstack{\textbf{Water Removal Factor:} \\Brain mask}}   
                                                            & \multirow{2}{*}{Mean} & \cellcolor{customRed!80}34.18 
                                                            & \cellcolor{customRed!80}53.09 & \cellcolor{customRed!80}224.20    \\ \cline{3-5}
                                                            &                       & \cellcolor{customYellow!80}18.84 
                                                            & \cellcolor{customYellow!80}18.53 & \cellcolor{customYellow!80}91.51 \\ \cline{2-5}
                                                            &\multirow{2}{*}{Median}& \cellcolor{customRed!80}17.19 
                                                            & \cellcolor{customRed!80}26.40 & \cellcolor{customRed!80}69.81     \\ \cline{3-5}
                                                            &                       & \cellcolor{customYellow!80}15.54 
                                                            & \cellcolor{customYellow!80}16.43 & \cellcolor{customYellow!80}12.54  \\ \hline
    \end{tabular}
    \caption{\rev{Lipid and Water Removal Factor Comparison. WALINET results are shown in \textcolor{customRed}{red}, HLSVD+L2 in \textcolor{customYellow}{yellow}. Lipid removal factors were computed in the head and brain, water removal factor was in the brain only. Results are shown for to subjects and on simulated data.}}
    \label{tab:lipTable}
\end{table*}

\end{document}